\documentclass{amsart}
\pdfoutput=1

\usepackage{graphicx}
\usepackage{amsmath,amssymb,amsfonts}
\usepackage[linesnumbered,ruled,vlined,norelsize]{algorithm2e}
\usepackage{xcolor}
\usepackage{overpic}
\usepackage{pgfplots}
\usepackage{pgfplotstable}
\pgfplotsset{compat=1.5}
\usepackage{array}
\usepackage{booktabs}

\def\R{\mathbb{R}}

\DeclareMathOperator*{\argmax}{arg\,max}

\newcommand{\mv}[1]{\mathbf{#1}}
\newcommand{\oldstuff}[1]{}

\begin{document}

\title{Iteratively reweighted greedy set cover}
\author{Marc Alexa}
\address{TU Berlin, Faculty of Computer Science\\Sekr. MAR 6-6, Marchstr. 23, 10587 Berlin}
\email{marc.alexa@tu-berlin.de}

\begin{abstract}
We empirically analyze a simple heuristic for large sparse set cover problems. It uses the weighted greedy algorithm as a basic building block. By multiplicative updates of the weights attached to the elements, the greedy solution is iteratively improved. The implementation of this algorithm is trivial and the algorithm is essentially free of parameters that would require tuning. More iterations can only improve the solution. This set of features makes the approach attractive for practical problems.
\end{abstract}


\maketitle

\section{Introduction}

The \emph{set cover problem} is one of the essential combinatorial tasks that are of theoretical as well as practical interest. Given are a finite number $n$ of elements, which we identify with the non-negative integers $\mathbb{Z}_n$ for notational convenience, and a finite number $m$ of sets
\begin{equation}
\mathcal{S}_i \subseteq \mathbb{Z}_n, \quad i \in \mathbb{Z}_m
\end{equation}
containing subsets of the elements.
A \emph{cover} is a collection $\mathcal{L} \subseteq \mathbb{Z}_m$ of the sets whose union contains all elements
\begin{equation}
\bigcup_{i \in \mathcal{L}} \mathcal{S}_i = \mathbb{Z}_n.
\end{equation}
The task is to find a \emph{minimal} cover, i.e.\ a collection among all covers so that no other collection covering the elements would consist of fewer sets or, if the sets have different \emph{cost}, so that no other cover would have a smaller sum of the costs. Finding such a minimal collection is long known to be NP-hard~\cite{Garey:1979}.

A large number of theoretical results are available on approximations to the set cover problem; the main motivation for this work is that the theoretical properties have been found to be poor predictors for practical performance~\cite{Grossman:1997}. We focus exclusively on practical aspects of unicost instances here -- there are no (new or known) theoretical results. Three commonly used basic principles for heuristics in this context are 1) stochastic search~\cite{Beasley1996,Ohlsson2001,Musliu2006}, 2) exploiting the linear programming solution to the fractional problem~\cite{Hochbaum1982,Peleg1993} (also using randomness~\cite{Peleg1997}) and 3) the greedy algorithm~\cite{Chvatal:1979} and variations derived from it potentially using the mentioned principles of randomization and/or the fractional LP solution~\cite{Grossman:1997,Haouari2002,Lan:2007}. 

Given the large number of existing heuristics, what is the aspect deserving further attention? We claim, similar to \cite{Lan:2007} that \emph{simplicity} is key for approximations to be useful in practice. By simplicity, we consider
\begin{enumerate}
\item the difficulty of implementing the approximation and
\item the necessity to choose parameters for the heuristic to perform well on a given instance.
\end{enumerate}
The greedy algorithm performs quite well on a range of different problems~\cite{Grossman:1997}, is easy to implement, and has no parameters in its basic version. In this work, we are using a version of the greedy algorithm that considers the elements to be weighted. This algorithm is used as a basic building block, and the weights are iteratively adjusted following the well-known multiplicative update strategy~\cite{Arora:2012:MWU}. The adjustment is extremely simple, has no parameters that require tuning, and it can only improve over time. We feel that makes the algorithm very attractive for practical deployment. 

We start by investigating several implementations of the greedy algorithm for large sparse problems (see Section~\ref{sec:greedy}). We observe that straightforward implementations of the greedy algorithm would suffer from the access cost of the usual sparse matrix representations. We show that it is advantageous to store the elements of each set \emph{and} the sets for each element, thus effectively doubling storage requirements. This makes it possible to treat much larger set cover problems with the greedy algorithm.

Given the implementation of the weighted greedy set cover, the question is how to adjust the per element weights to reduce the number of sets. The basic idea is based on work by \cite{Broennimann1995}, namely by attempting a $k$-cover for considerably small $k$, and increasing weights of elements that were left uncovered with $k$ sets. Our main contribution could be considered the specific scheme for adjusting $k$ and the weights (see Section~\ref{sec:iteration}), based on experimental results across different instances. Our approach has only one parameter: the multiplicative growth factor for the weight of one uncovered element.

In Section~\ref{sec:experiments} we apply the algorithm to some known instances of the set cover problem. In a large scale experiment we show that a wide range of factors for increasing the weights work well on different types of problems. We also show how on many problems  the algorithm improves the greedy solution to roughly 70\% of the initially covering sets. The instance we are particularly interested in is a geometric set cover problem, which we briefly introduce in the Appendix.

We conclude in Section~\ref{sec:conclusions}, in particular with an interpretation of the solution in a more general way. This suggest some avenues for further exploring the idea, including how the approach could be used for non-unicost problems. 

\section{Sparse weighted greedy set cover}
\label{sec:greedy}

The main idea of the greedy algorithm~\cite{Chvatal:1979} is to pick, in each step, the set with the highest \emph{value}. If all elements are treated equally, the value of a set is simply the number of elements that would be \emph{additionally} covered if the set was chosen, i.e.\ the number of elements it contains, not already covered by a set that has been selected earlier. Note that the algorithm can be used for both, computing a covering set or computing a collection of $k$ sets that cover many elements.  

We need a weighted version of this algorithm, in the sense of the elements carrying weights (and not the sets). Let $w_i$ be the weight of element $i$.
A common way to think of set cover problems is by representing the sets in form of a matrix: each row corresponds to an element, each column represents a set. Let 
\begin{equation}
V = \{ v_{ij} \},\quad v_{ij} = \begin{cases} w_i & j \in \mathcal{S}_j\\ 0 & \mathrm{else} \end{cases}
\end{equation}
be this matrix, $\mathbf{v}_i$ its $i$-th column, and $\mathbf{v}^j$ its $j$-th row. Initially, the value of the set with index $i$ is $\| \mathbf{v}_i \|_1$. 

In each step the column with maximal value $\argmax_i \| \mathbf{v}_i \|_1$ is selected. Then all columns for which $\mathbf{v}_i$ has non-zero entries are deleted; or, equivalently, all elements $j \in \mathcal{S}_i$ are excluded. With deletion of the columns the values of the sets change. Note that the $i$-th column need not be deleted from the matrix, because its value will be zero after deletion of the rows. The steps are repeated until all rows have been removed. 

The important observation for sparse instances is that the removal of elements (rows) in a step affects the value of only very few sets (columns). In other words, it is wasteful to recompute the values of all sets -- only few sets require an update of their value. The sets are identified by having non-empty intersection with the selected set $\mathcal{S}_i$, i.e.\ for each element $j \in \mathcal{S}_i$ we need to find the sets $\mathcal{S}_k$ that contain $j$. In terms of the matrix representation this means we need to identify the non-zero elements in column $i$, then remove the corresponding rows, and finally update the sums of only the affected columns. 

In other words, an implementation that avoids accessing all elements requires access to the non-zero elements of columns and rows. The problem is that sparse matrices are commonly stored in either \emph{column major} or \emph{row major}, which corresponds to representing either the sets (i.e.\ storing the $\mathcal{S}_i$ as vectors) or storing for each element the sets that contain the element. This storage format makes the access to either the elements of a set or the sets containing an element inefficient.  

For this reason we suggest to store coverage information in ``both directions'': store the elements of the sets $\mathcal{S}_i$ and also the \emph{coverage} information
\begin{equation}
\mathcal{C}_j = \{i: j \in \mathcal{S}_i\}.
\end{equation}
Note that each collection of sets can be generated in optimal time complexity from the other collection of sets (i.e.\ constant time per set-element relationship). 

We have implemented three versions of the greedy set cover algorithm in C++ to demonstrate the advantage of storing the covering relationships both ways. As sparse matrix library we use Eigen~\cite{eigenweb}, which is known to be very good in exploiting modern hardware. 
\begin{enumerate}
\item An implementation based on a row major sparse matrix representation (it is rather obvious that column major would be inefficient).
\item An implementation creating both row and column major representations. 
\item An implementation based on simply representing the covering relationships $\mathcal{S}$ and $\mathcal{S}$ as arrays. 
\end{enumerate}
An important detail in all implementations is to never actually remove elements (or rows) but to keep a binary array that flags them as inactive. Rows (or elements) that have been covered by a selected set are then kipped. This turned out to be significantly more efficient.

\begin{table*}
\begin{tabular}{ >{\footnotesize \sc} l | r | r | r | r | r | r | r | r | r | r }
  \hline                       
  \normalsize{\normalfont{Source}} & $n$ & $m$  & $\frac{\sum |\mathcal{S}_i|}{nm}$  & k & \multicolumn{2}{c|}{row major}   & \multicolumn{2}{c|}{col+row}  & \multicolumn{2}{c}{2 arrays} \\
        &    & & &  & pre  & gsc & pre  & gsc & pre & gsc\\
  \hline
scpcyc11 & 28160 & 11264 & 0.04\% & 4304 & 0.01 & 0.71 & 0.01 & 0.09 & 0.01 & 0.16\\
Bunny 1\% & 34834 & 34768 & 0.73\% & 225 & 0.69 & 0.23 & 1.08 & 0.06 & 0.25 & 0.04\\
Bunny 2\% & 34834 & 34768 & 1.5\% & 88 & 1.49 & 0.24 & 2.32 & 0.09 & 0.53 & 0.05\\
Bunny 4\% & 34834 & 34768 &  2.4\% & 36 & 3.52 & 0.28 & 5.91 & 0.17 & 1.33 & 0.10\\
VLion .5\%& 200K & 199K & 0.10\% & 1920 & 3.00 & 10.8 & 4.82 & 0.89 & 1.00 & 0.69\\
VLion 1\%& 200K & 199K & 0.24\% & 665 & 7.55 & 6.76 & 12.1 & 0.62 & 2.38 & 0.45\\
VLion 2\%& 200K & 199K & 0.63\% & 160 & 26.5 & 4.51 & 42.9 & 1.04 & 8.00 & 0.68\\
Vase .5\% & 896K & 688K & 0.07\% & 2944 & 31.8 & 260 & 59.1 & 6.11 & 8.27 & 3.72\\
Vase 1\% & 896K & 688K & 0.16\% & 1314 & 149 & 285 & 317 & 8.58 & 23.1 & 3.39\\
Vase 2\% & 896K & 688K & 0.40\% & 546 & - & - & - & - & 120 & 32.3\\
\end{tabular}
\label{tab:results}
\end{table*}
We compare the implementations on several large set cover problems -- one of them from the publicly available set at OR, and several significantly larger ones based on our application scenario (see Appendix 1). The statistics of this comparison are shown in Table~\ref{tab:results}. Timings are runtimes in seconds and should be taken only as rough indicators of relative performance.  The numbers show that storing both covering relationships is clearly beneficial. The sparse matrix representation actually incurs some overhead that we avoid by directly storing the covering relationships.  

For each algorithm we provide the time for setting up the data structures (sparse matrices) and the time for generating the cover once the data structures are in place. For the heuristic we explain in the following section we need to repeatedly create covers, while the data structures need to be created only once. In this context, our implementation is two order of magnitude faster than the implementation based on a row major sparse matrix representation. Moreover, for the largest problem we tried (which has roughly a billion covering relationships), the matrix version was aborted by the kernel due to excessive use of resources.

Our implementation of the greedy set cover is straightforward. We provide a version that considers weighted elements and stops after either all elements are covered or at most $k$ sets have been used. Pseudocode is given below. We also provide a single C++ file containing the implementation of all three implementations and generates the relevant statistics online.

\begin{procedure}
\DontPrintSemicolon
\SetKwInOut{Data}{data}\SetKwInOut{Input}{input}\SetKwInOut{Output}{output}
\Data{Covering relationships $\mathcal{C},\mathcal{S}$}
\Input{Number $k$ of covering sets, weights $\{w_j\}$}
\Output{Binary arrays $\mv{c} \in \{0,1\}^n, \mv{s} \in \{0,1\}^m$ of selected sets and covered elements}
$\mv{c} \leftarrow \{0\}^n, \mv{s} \leftarrow \{0\}^m$\;
\For {$i \in [0,\ldots,n-1]$} {$v_i = \sum_{j \in \mathcal{S}_i} w_j$\;}
\While {$\|\mv{s}\|_0 < k \cap \| \mv{c} \|_0 < n$}
{
	$i \leftarrow \argmax_i u_i$\;
	$s_i \leftarrow 1$\;
	\For{$j \in \mathcal{S}_i$}
	{
		\If{$c_j = 0$}
		{
			$c_j \leftarrow 1$\;
			\For{$l \in \mathcal{C}_j$}
			{
				$u_l \leftarrow u_l - w_j$\;
			}
		}
	}
}
\caption{$k$-cover()}
\label{alg:greedy-k-cover}
\end{procedure}

\section{Iterative reweighting heuristic}
\label{sec:iteration}

A central idea of Br{\"{o}}nnimann and Goodrich~\cite{Broennimann1995} is to check if a $k$-cover exists for fixed $k$ by repeatedly running the weighted $k$-cover algorithm and increase the weights of elements that were left uncovered. This idea is part of a constructive proof, and the details of how to guess $k$ and in particular how to increase the weights for which elements are insignificant for the asymptotic complexity analysis and consequently not discussed. 

Our basic strategy is to start by running the greedy algorithm to generate an initial $k$-cover. Then we try to find a $k-1$ cover by re-weighting. If successful, we continue with $k-2$. We can do this as long as the user is willing to wait. This strategy has two important features:
\begin{enumerate}
\item The heuristic can be run on any time budget. More time can only improve the result -- never make it worse.
\item It is useful to keep the weights unchanged when $k$ is reduced, as the weighting that led to the $k$-cover is often closer to a weighting allowing a $k-1$ cover compared to uniform or random weighting.
\end{enumerate}

Through experimentation we have found that increasing the weight of only \emph{one} uncovered element is better than increasing the weights of all uncovered elements and, moreover, that multiplicative increments are better than additive ones -- the latter is in line with the popular multiplicative weight update strategy~\cite{Arora:2012:MWU}. We also found that the particular missing element chosen makes no significant difference. This allows us to take the most efficient solution and increase the weight of the first uncovered element encountered. With this approach in place, our heuristic has only a single parameter: the multiplicative factor for increasing the weight of the first uncovered element encountered. In the following section we demonstrate that the choice of this parameter is not critical for the instances we tested. 

Putting the ideas together leads to the algorithm detailed in the following pseudo code. We also provide the algorithm implemented in C++ on URL for further experimentation. 

\begin{procedure}
\DontPrintSemicolon
\SetKwInOut{Data}{data}\SetKwInOut{Input}{input}\SetKwInOut{Output}{output}
\Data{Growth factor $f$}
\Input{Covering relationships $\mathcal{C},\mathcal{S}$}
\Output{Binary array $\mv{s}^* \in \{0,1\}^m$ of selected sets}

$k \leftarrow m, \mv{w} \leftarrow \mv{1}^n$\;
\While {not stopped by user}
{
	$\mv{c},\mv{s}$ = $k$-cover($k,\mv{w}$)\;
	$i \leftarrow 0$\;
	\While{$i < n \wedge c_i = 1$} { $i \leftarrow i+1$\; }
	\eIf {$i < n$}
	{
		$w_i \leftarrow w_if$\;
	}
	{
		$\mv{s}^* \leftarrow \mv{s}$\;
		$k \leftarrow k - 1$\;
	}
}
\caption{Iterative reweighting for greedy set cover()}
\label{alg:reweighting}
\end{procedure}

The algorithm repeatedly searches for the first uncovered element and provides its index as $i$. If an index exists ($i<n$) the weight of the corresponding element is increased. Otherwise, a cover has been found. This cover is stored in $\mv{s}^*$ and $k$ is decremented.

The algorithm only depends on the choice of growth factor $f$. The solution(s) it provides also depend on the ordering of elements and sets. For one, also the greedy algorithm is dependent on the order of sets, as this defines how ties are broken among sets of equal value. Our heuristic also depends on the order of elements, as this defines which weight is increased. 

\begin{figure}
\centering
\includegraphics[width=0.475\textwidth]{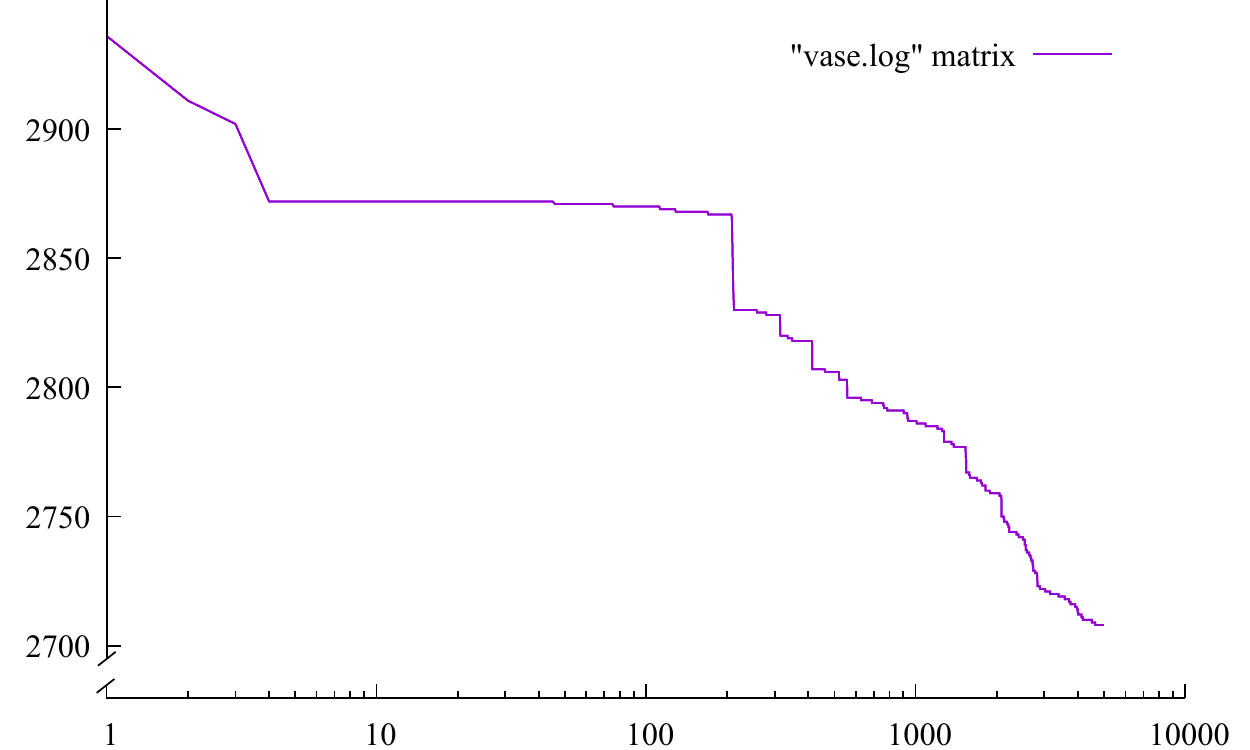}
\includegraphics[width=0.475\textwidth]{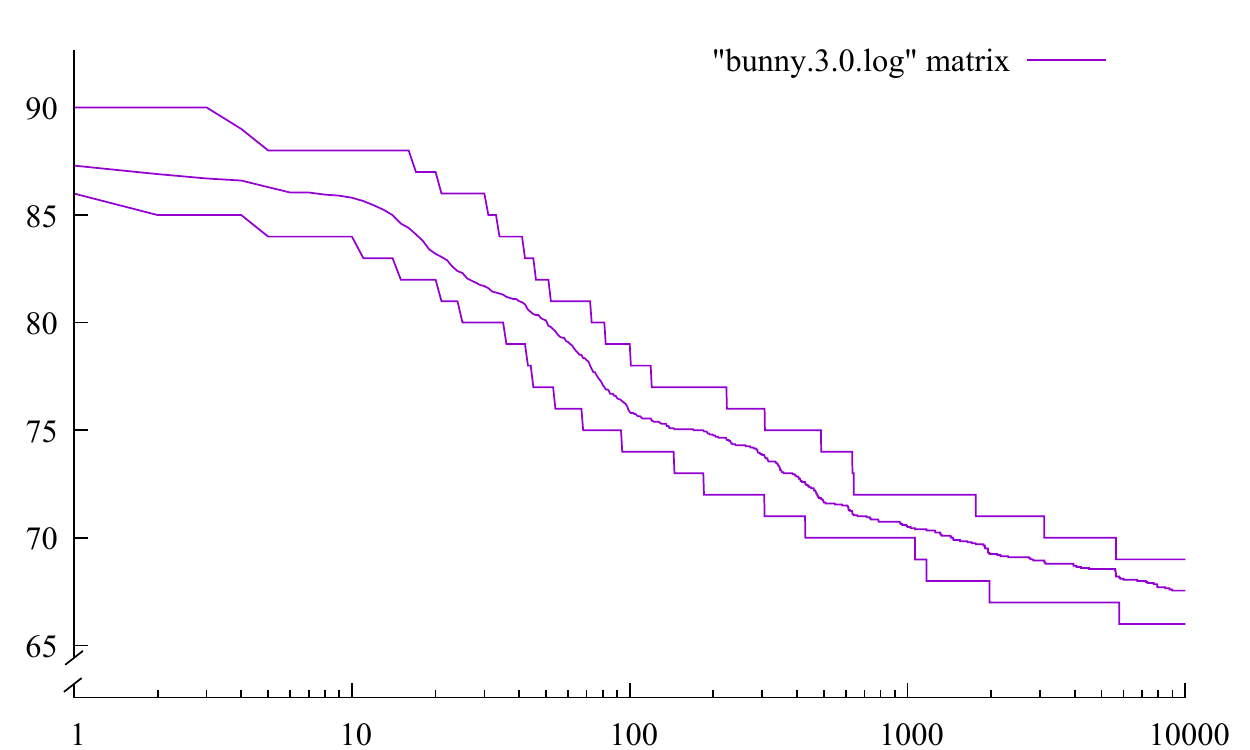}
\caption{The number of sets $k$ as a function of the iterations. Left: a single run of the algorithm on the \textsc{Vase} data (900K elements, 700K sets). Right: The minimum, mean, and maximum $k$ for permutations of the set order for the \textsc{Bunny} data (35K elements, 35K sets).}
\label{fig:kvsiter}
\end{figure}

\section{Experiments and results}
\label{sec:experiments}

We have applied the heuristic to a variety of test problems to gather some insight into the behavior and performance of the heuristic. The decrease of $k$ as a function of the iterations is shown for a single run of the algorithm on the \textsc{Vase} data in Figure~\ref{fig:kvsiter} (left). However, the performance of the algorithm depends on how ties are broken in the greedy algorithm, which in turn depends on the order of the set. So for analysis we run the algorithm several times on the same problem while randomly permuting the order of the sets. Then we can inspect the minimal, maximal and most importantly the mean value for $k$ as a function of the iterations. The right side of Figure~\ref{fig:kvsiter} provides one such graph for the \textsc{Bunny} data. 

Using the mean $k$, we first analyze the dependence on the growth factor, then compare to simple randomization of the greedy algorithm, and eventually show performance data for various test problems.

\subsection{Dependence on growth factor}

\begin{figure}
\centering
\minipage{0.66\textwidth}
\includegraphics[width=\linewidth]{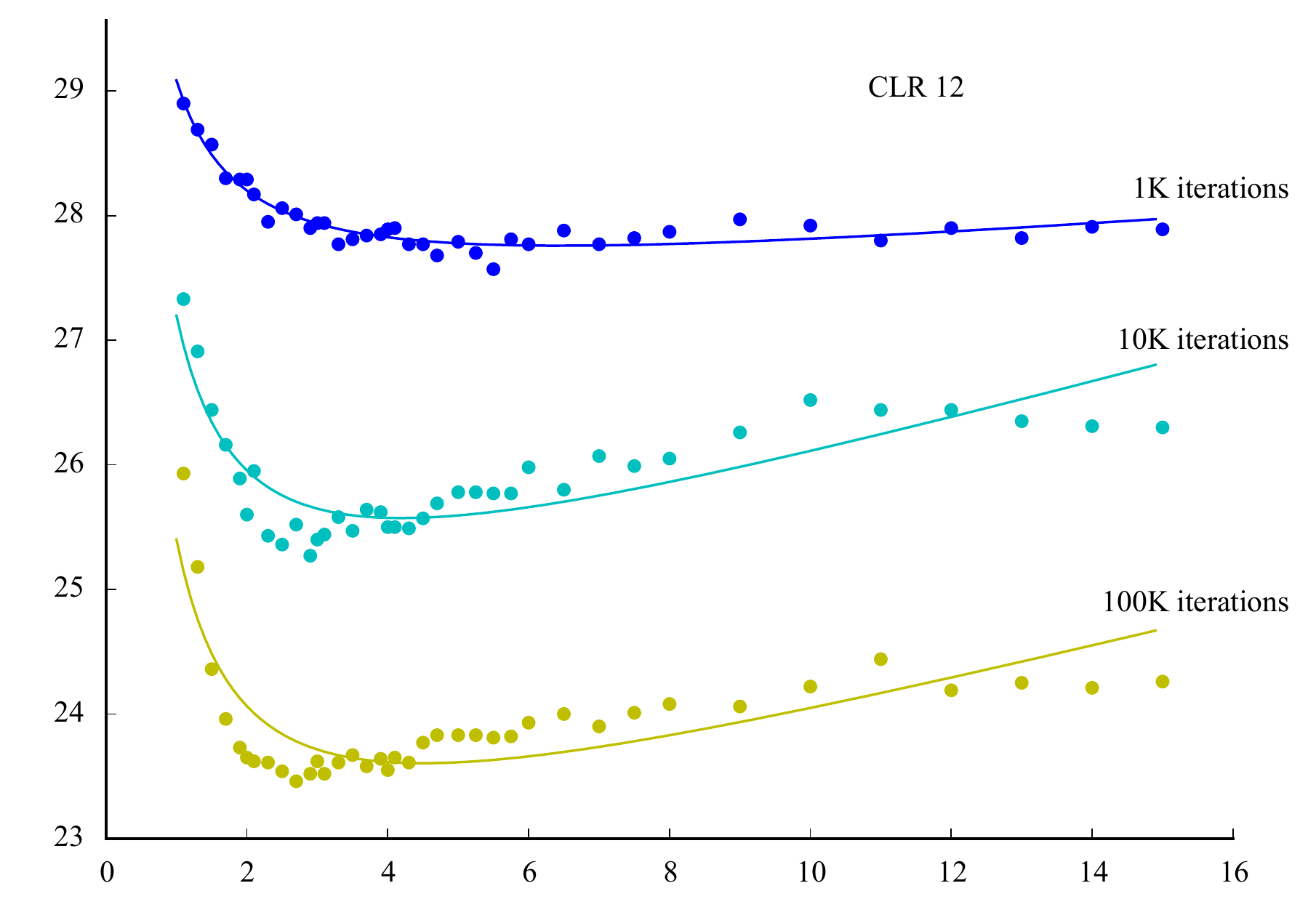}
\endminipage\hfill
\minipage{0.33\textwidth}
\includegraphics[width=\linewidth]{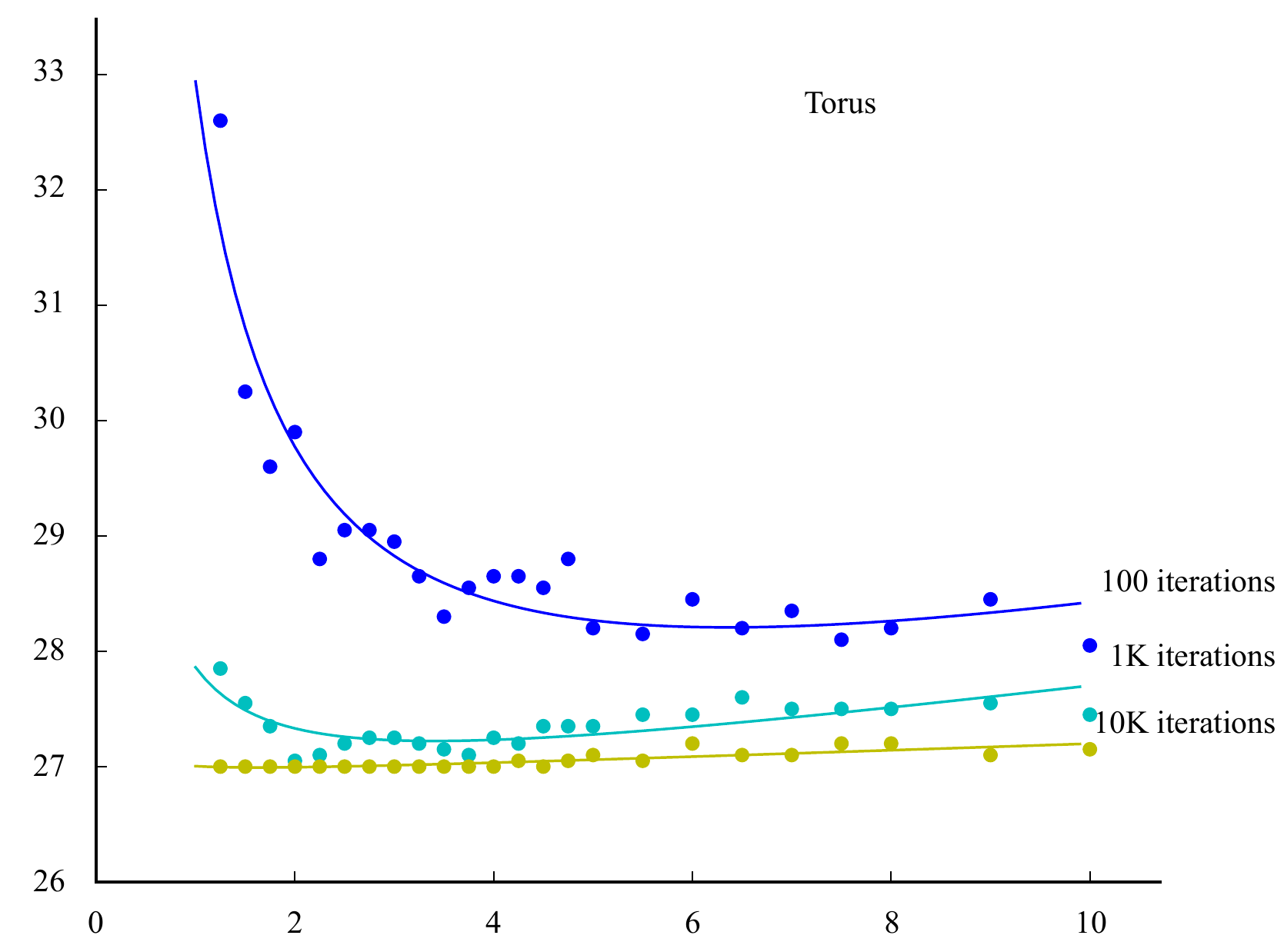}\\
\includegraphics[width=\linewidth]{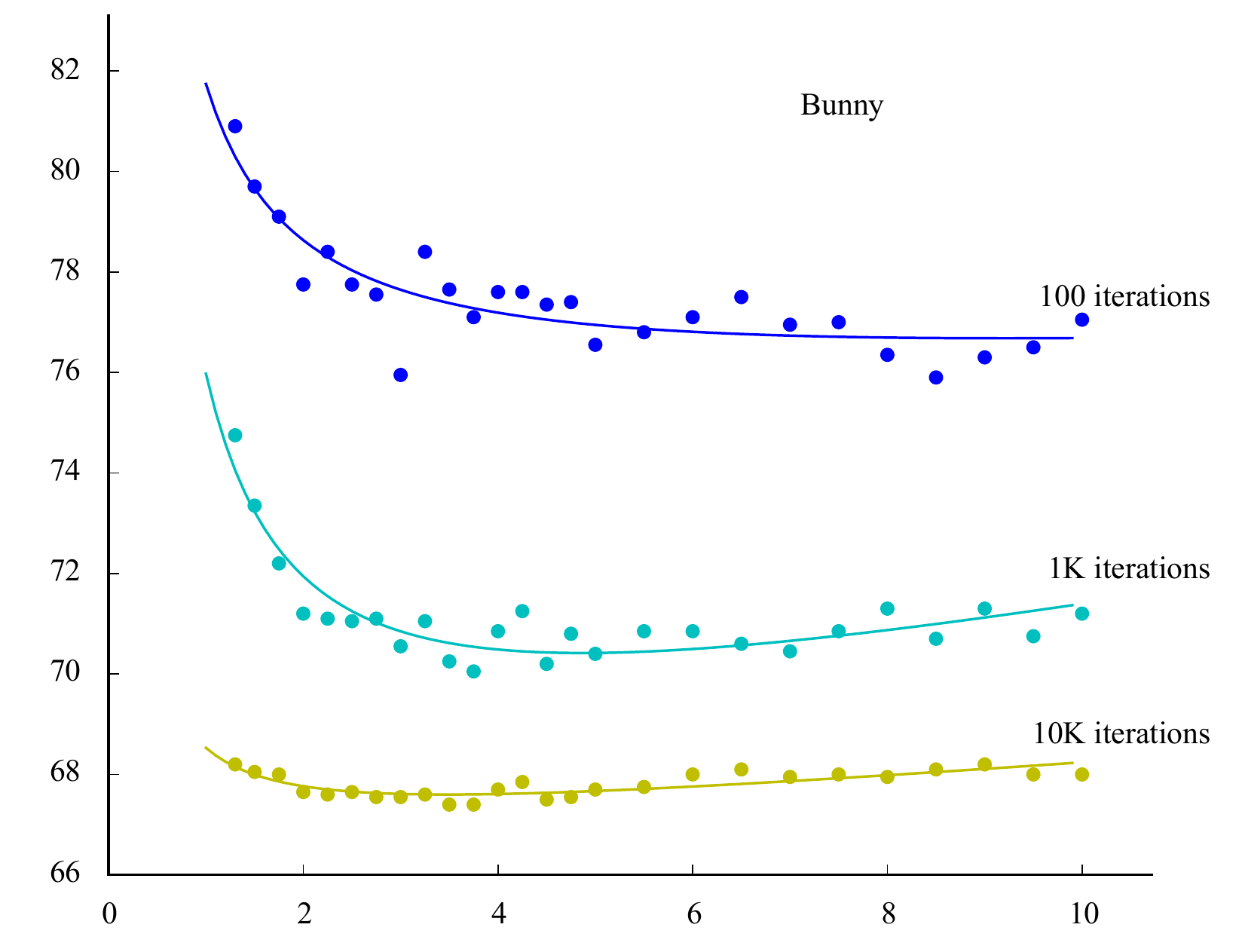}
\endminipage
\caption{The mean $k$ after a fixed number of iterations as a function of the growth parameter. Left: the CLR 12 data from the OR-library after 1K, 10K, and 100K iterations. Larger growth factors help to decrease $k$ quickly, but in the long run smaller growth factors generate better results. Right: two geometric instances with regular structure and 13K sets and elements (top) and 35K elements and sets but more irregular (bottom) after 100, 1K, 10K iterations. After sufficient number of iterations the dependence on the growth factor is minimal.}
\label{fig:growth}
\end{figure}

We selected three problems for sampling the performance of reweighting as it depends on the growth factor: the CLR.12 problem as given in the OR-library~\cite{Beasley1990}; and two instances from our geometric set cover problems, one based on a torus (13K elements and sets), exhibiting a lot a regularity, and one based on the \textsc{Bunny} data set ($n = 34834, m = 34768, \rho = 1.5\%$, see Appendix for more information on how the data was generated).

Rather than comparing the graphs of $k$ as a function of the iterations, we fix the number of iterations, and plot the mean $k$ as a function of the growth factor. For the CLR.12 data we chose iteration counts $(1K,10K,100K)$ and for the other two problems we used $(100,1K,10K)$ as they converge quicker. The resulting scatter plots are shown in Figure~\ref{fig:growth}.

The CLR.12 problems converge slowest among the three. Larger growth factors help in decreasing $k$ quickly in the early iterations, yet they appear to slow down convergence in the long run. The optimal solution $k=23$ is found for a wide range of growth factors after a sufficient number of iterations. It seems that values in the range $[2,5]$ for the growth factor are good choices for finding the optimum solution quickly.

On the geometric instances the choice of the growth factor appears even less critical. For a very wide range of values the optimization generates similar solutions.

We have settled for a growth factor of 3 for all our following experiments. It is unlikely that a slightly different choice would lead to fundamentally different results for the problems we considered.

\subsection{Random permutations}

With the growth factor being fixed, the main feature of reweighting is simplicity. The algorithm is only marginally more complex than the greedy algorithm itself and there are parameters left that would require (or offer) tuning. One of the few approaches that is arguably even simpler is searching among the \emph{different} unweighted solutions of the greedy algorithm, where the difference comes from how ties are broken. We already perform this randomization in our experiments by permuting the order of the sets.

It is known that this randomization fails to systematically improve the performance of the greedy algorithm~\cite{Grossman:1997}. Since our approach does the question is: how many iterations are necessary so that reweighting outperforms random sampling? 

\begin{figure}
\centering
\includegraphics[width=0.475\textwidth]{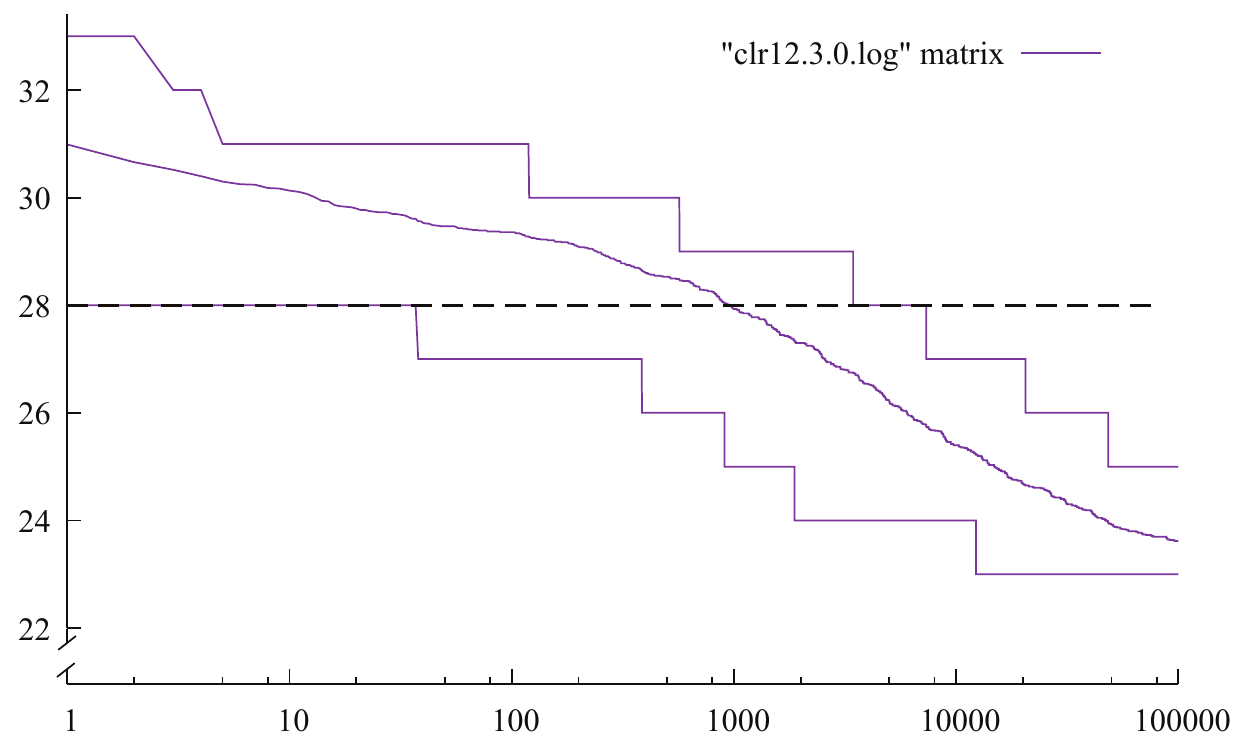}
\includegraphics[width=0.475\textwidth]{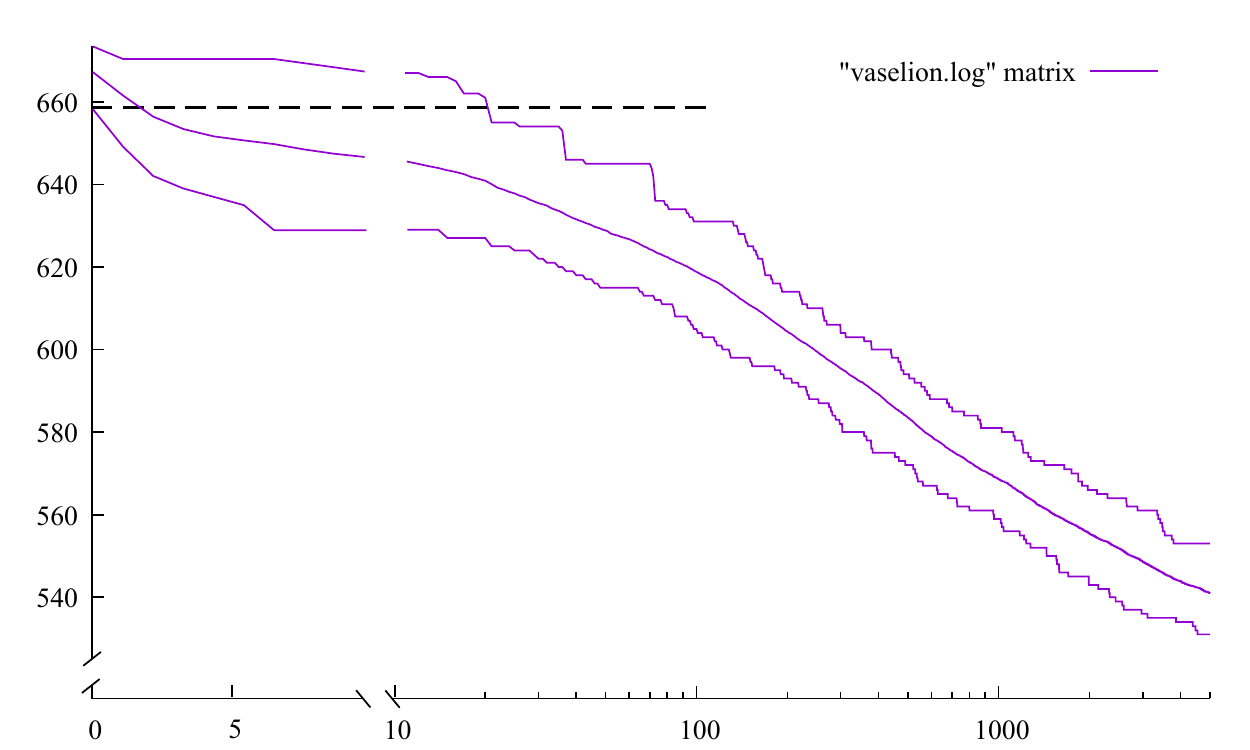}
\caption{Performance of reweighting relative to permutations. The performance of permuting the order of elements in the greedy algorithm is represented by the gap between the minimum and maximum $k$ after 0 iterations (i.e.\ at the intersection of the graphs with the vertical axis). Left: the best greedy solution among $10^6$ permutations for the CLR12 data was $k=28$. Reweighting outperforms this solution on average after roughly 1000 iterations (intersection with the mean curve, and always after roughly 7000 iterations (intersection with the max curve). Right: on large sparse instance the result is more pronounced. For the \textsc{Vaselion} data (200K elements and sets) the best solution among $10^5$ permutations is outperformed in the mean after 2 and in the worst case after 20 iterations.}
\label{fig:perm}
\end{figure}

For CLR.12 data, we find a minimum of $k = 28$ from about 100 permutations of the sets. We have increased the number of permutations to $10^6$, but no better value has been found. As can be seen in Figure~\ref{fig:perm}, for about $1K$ iterations of reweighing the mean $k$ is systematically lower than 28. At around $7K$ iterations all 100 random trials we performed were at $k=27$, meaning even the worst case performance outperformed the simple random permutation of the greedy algorithm. 

The geometric instances profit even less from random perturbations and more from reweighting. This could have already been deduced for the \textsc{Bunny} data illustrated in Figure~\ref{fig:kvsiter}. We analyzed this in more detail for the larger \textsc{Vaselion} data (200K elements and sets). In $10^5$ permutations the smallest cover used $k = 657$ sets. Over 100 permutations, the mean solution was better after only 2 iterations, and all 100 trials were better after 21 iterations, see Figure~\ref{fig:perm}.

\subsection{Performance}

\begin{figure}
\centering
\includegraphics[width=0.6\textwidth]{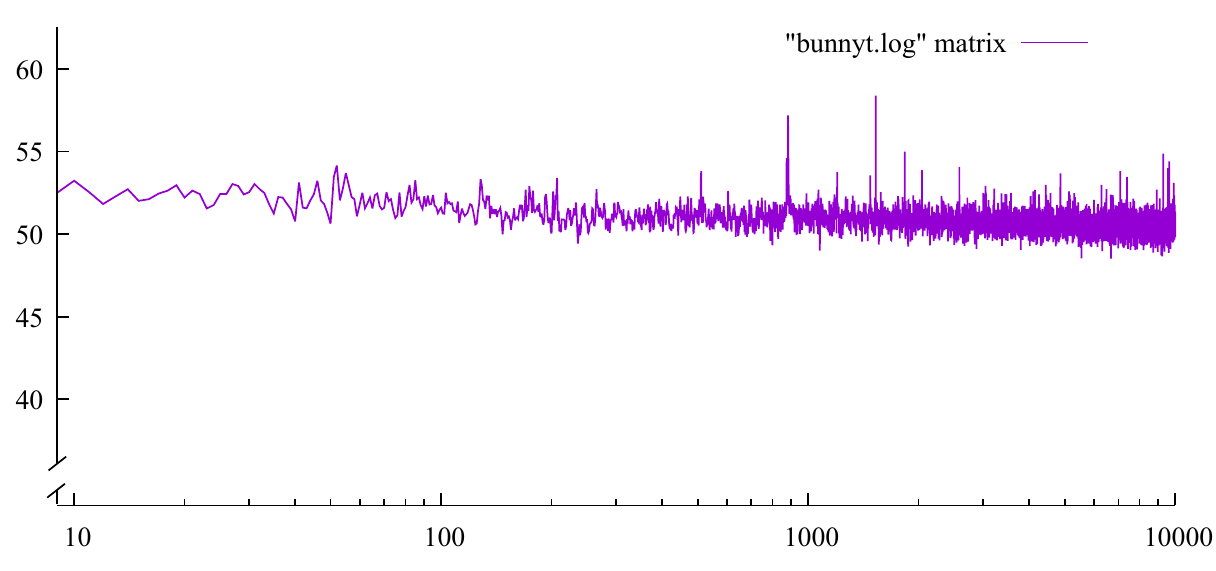}
\hfill
\includegraphics[width=0.3\textwidth]{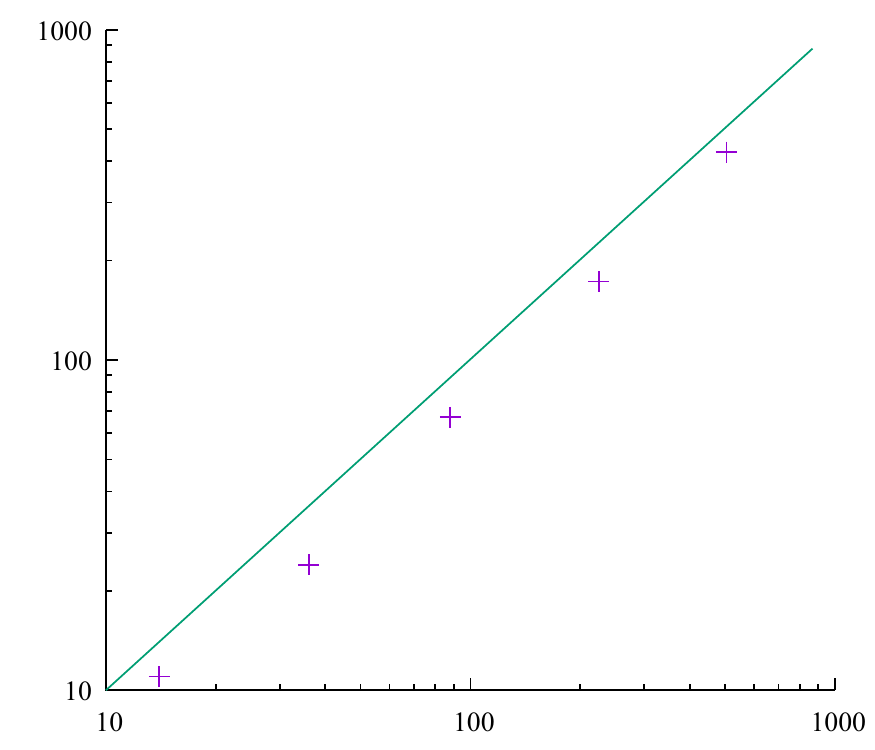}
\caption{Left: Time for weighted greedy solve against iterations for the \textsc{Bunny} data. The times are stable and slightly decrease (due to $k$ reducing). Right: Improvements from reweighting are stable for different densities. The data for the \textsc{Bunny} with 35K elements and sets has been generated with different densities, i.e.\ different numbers of elements per set. The resulting covers have different $k$. The plot shows $k$ of the greedy algorithm on the horizontal axis against $k$ from reweighting on the vertical axis. The distance to the bisector in this log-log plot indicates relative improvement over the greedy solution.}
\label{fig:stats}
\end{figure}

We are not aware of other heuristics that have been tested on the size of problems we are interested in. On the unicost problems in the OR-library~\cite{Beasley1990} our approach appears to perform roughly similar to other successful heuristics~\cite{Musliu2006,Lan:2007}. In particular, it finds the optimal solution for the smaller CLR problems (like the other heuristics do) and provides approximations for the CYC problems that are roughly similar the data provided by \cite{Lan:2007} but slightly worse than what is reported by \cite{Musliu2006}. We want to stress that comparing the best solutions to particular instances may be misleading as a predictor of the practical value of a heuristic: timing often depends on a variety of factors, worst case behavior is commonly not reported, and in many case parameter tuning was necessary and is only scarcely documented. In comparison, there is no adjustment of parameters in our approach. The time spent for the computation depends on the number of iterations one is willing to invest and each iterations requires the amount of time for the greedy algorithm --- Figure~\ref{fig:stats} shows that the variation of time needed for each iteration is minor. 

Our reweighting scheme consistently shows significant improvement over the greedy solution for large sparse problems up to a million elements and million sets (see graphs in Figures~\ref{fig:kvsiter} and \ref{fig:perm}). The effect is also stable across different densities. For the \textsc{Bunny} we generate sets of different densities, such that in all cases the number of elements and sets are the same (35K). Because of the different densities the number of required sets $k$ is different, but the improvement of the reweighting for $10^5$ iterations over the initial greedy solution is similar (see the log-log plot in Figure~\ref{fig:stats}).

\section{Conclusions and future work}
\label{sec:conclusions}

We have shown how to implement the greedy algorithm for the set cover problem for large sparse data sets. Reweighting the elements leads to an algorithm that has the desirable properties of straightforward  implementation and no parameter tuning.

It would be possible to combine this algorithm with meta-heuristics that have proved successful for set cover (e.g.\ tabu search). This would likely lead to better covers for some problems, albeit at the expense of making the algorithm more difficult to implement efficiently and introducing parameters that require adjustment to the characteristics of the data. We suggest this avenue only if the smallest cover for a particular instance is of interest.

While we have analyzed the approach for the unicost problem, it works exactly the same if the costs of the sets are different: the greedy algorithm is usually set up to maximize the number of uncovered elements divided by the cost of the set~\cite{Chvatal:1979}, and still performs well on a variety of problems. Reweighting will work as for the unicost problems. As the problems we are concerned with are unicost, we decided to leave the experimental analysis of other situations as future work. 

Lastly, let us interpret the given approach out of the perspective parameterizations of the set cover problem. The dominant way to search the space of solutions is by considering the chosen sets. This could be interpreted as a binary vector of length $m$, the number of sets. One may instead parameterize the current solution using a real valued vector of length $n$; and the greedy algorithm for weighted elements serves as a device for translating the vertex weights into the selected sets. The neighborhood structures of the two parameterization are likely very different. We have presented one way of exploring the neighborhood --- it would be interesting to apply other strategies for searching the space of element weights. 

\bibliographystyle{amsplain}
\bibliography{scp}


\appendix
\section{Covering discrete surfaces with enlarged medial balls}

The background for this work is as follows: given points $\mv{p}_i \in \R^3$ representing the surface of a solid body, we wish to approximate the shape with a set of balls $\mathcal B_j = (\mv{c}_j,r_j)$. We say a point $i$ is contained in ball $j$, i.e.\ $\mv{p}_i \in \mathcal B_j$, if $\| \mv{p}_i-\mv{c}_j\| < r_j$. We wish to find a small number of balls whose union cover the whole point set, i.e. $\{\mv{p}_i\} \subseteq \cup_j \mathcal B_j$, while at the same time the union of the interior of the balls is not much larger than the interior of the shape. 

This problem can be solved by generating a set of balls whose interior has a small distance to the interior of the shape and then finding a small subset that covers all points --- this is the set cover problem we are interested in. It makes sense to construct the set of balls by increasing the radius of \emph{maximal empty interior} balls. By empty we mean that ball contains none of the $\mv{p}_i$ in its interior, i.e.\ $\mathcal B_j \cap \{\mv{p}_i\} = \emptyset$. By maximal we mean that any ball $\mathcal B'_j$ that strictly contains $\mathcal B_j$ also contains a point in the point set. The idea is that any ball in the cover that is contained in a larger ball will never be needed.  Any such empty maximal ball will have a Voronoi vertex of the point set as its center~\cite{Amenta:1998:SRV:276884.276889}. Lastly, by interior we mean that the center of the ball is in the interior of the shape. There are usually more interior Voronoi vertices than samples and it is common to consider only the largest ball for each $\mv{p}_i$~\cite{Amenta:1998:NVS:280814.280947}.

Increasing the radius of the ball by a fixed amount $\epsilon$ will induce a set difference between the shape and the ball of at most $\epsilon$. Thus, also the union of the enlarged balls (and also any subset) has bounded set distance to the interior. This means solving the set cover problem provides solutions that minimize the number of balls while guaranteeing a certain error. Varying $\epsilon$ sets up set cover problems with varying density, as the number of elements to be covered (the points) as well as the number of sets (the balls) are (largely) independent of $\epsilon$. Yet, because surface samples describing shapes are large, the resulting set cover problems are large.

\begin{figure}
\centering
\includegraphics[width=\textwidth]{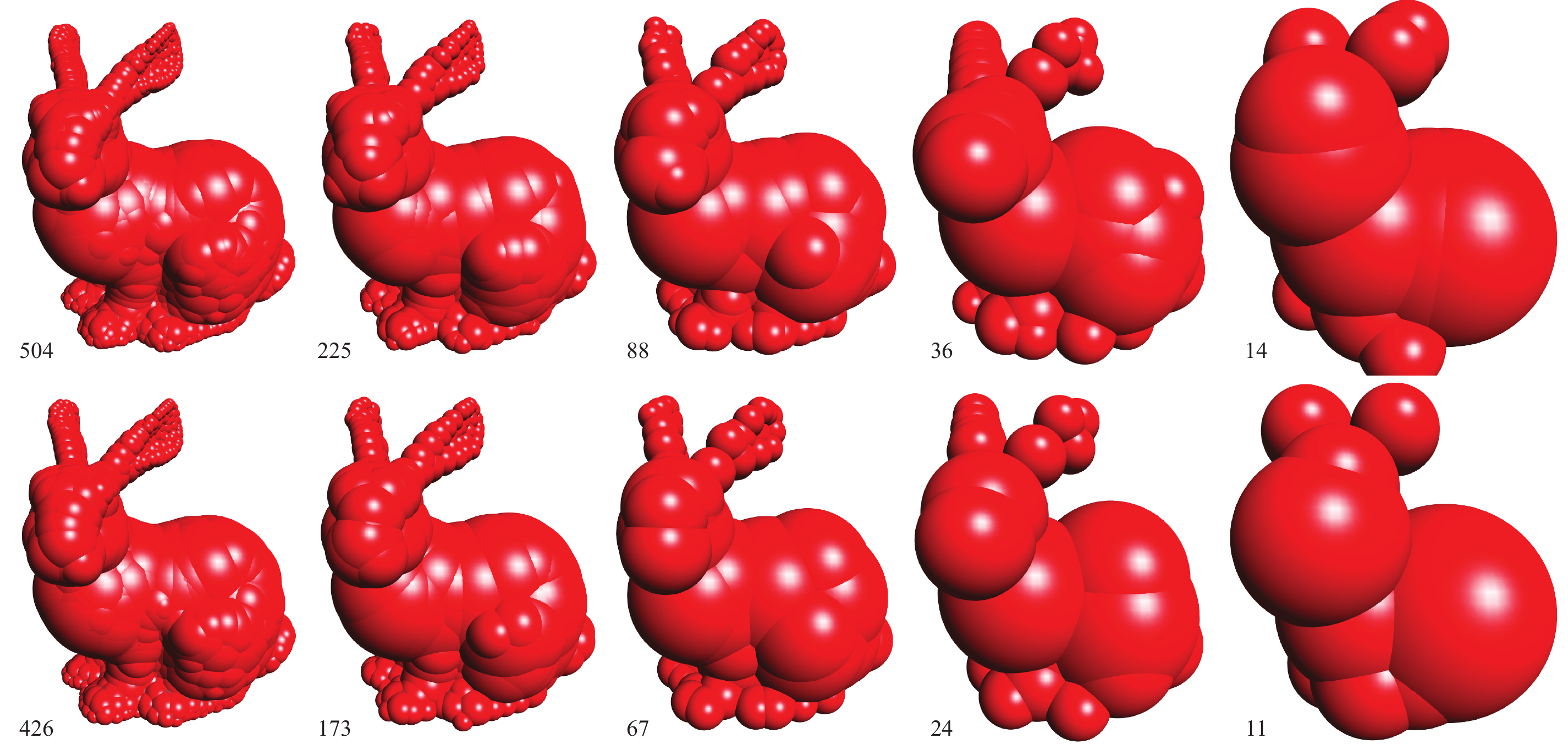}
\caption{Optimizing the greedy set cover result (top row) significantly reduces the number of spheres to cover the shape (bottom row), while satisfying the same bound on the approximation error.}
\label{fig:coveropt}
\end{figure}

Figure~\ref{fig:coveropt} provides examples of the covers obtained with algorithms presented here. The empty balls have been enlarged by 0.5\%, 1\%, 2\%, 4\%, 8\% of the bounding box diagonal of the shape. The top row shows the covers resulting from applying the greedy algorithm, the bottom row the improved covers using the reweighting strategy. 

\end{document}